\documentclass{svproc}
\usepackage{url}

\usepackage{amsmath}
\usepackage{amssymb}
\usepackage{float}
\usepackage{graphicx}
\usepackage{cleveref}
\usepackage{color}
\usepackage{subcaption}

\def\Var{{\textrm{Var}}\,}
\def\E{{\mathbb{E}}\,}
\def\Cov{{\textrm{Cov}}\,}

\begin{document}
\mainmatter              % start of a contribution
\title{Using a Bayesian approach to reconstruct graph statistics after edge sampling}
\titlerunning{Sampled graph reconstruction}  % abbreviated title (for running head)
%                                     also used for the TOC unless
%                                     \toctitle is used
%
\author{Naomi A. Arnold \and Ra\'ul J. Mondrag\'on \and Richard G. Clegg}
\authorrunning{Naomi A. Arnold et al.} % abbreviated author list (for running head)
%
%%%% list of authors for the TOC (use if author list has to be modified)
\tocauthor{Naomi A. Arnold, Raul J. Mondrag\'on and Richard G. Clegg}
\institute{Queen Mary University of London, London, E1 4NS, UK\\
\email{naomi.arnold@nulondon.ac.uk}}

\maketitle              % typeset the title of the contribution

\begin{abstract}
    Often, due to prohibitively large size or to limits to data collecting APIs, it is not possible to work with a complete network dataset and sampling is required. A type of sampling which is consistent with Twitter API restrictions is uniform edge sampling. In this paper, we propose a methodology for the recovery of two fundamental network properties from an edge-sampled network: the degree distribution and the triangle count (we estimate the totals for the network and the counts associated with each edge). We use a Bayesian approach and show a range of methods for constructing a prior which does not require assumptions about the original network. Our approach is tested on two synthetic and three real datasets with diverse sizes, degree distributions, degree-degree correlations and triangle count distributions.
\end{abstract}

\section{Introduction}\label{sec:intro}Analysis of complex networks remains a growing area and network data sets are more and more commonly available. However, some data sets are only a sample of the entire network. For very large networks, it may not be possible to work with complete data because of its size.
%The degree distribution and triangle count are two simple yet fundamental descriptors of a network. 
%Thanks to the APIs and the rise in popularity of open decentralised systems such as the blockchain, there is more network data than ever. 
%However, it is not always possible to obtain statistics like the degree and triangle distributions for the whole network. 
Additionally, APIs can rate-limit the number of queries, meaning that not all nodes and edges are present~\cite{morstatter2013sample}. A common example is the Twitter stream API which returns a 1\% random sample of all tweets in real-time~\cite{twitter2022sampling}. In the usual Twitter graph formulation where edges constitute 1:1 replies or retweets, this corresponds to uniform edge sampling of the full Twitter reply/retweet graph. Inferring even simple characteristics such as the true number of nodes or edges from a sample can be nontrivial~\cite{katzir2011estimating,bianconi2022grand}. 
%[citation needed]
% I don't think we need a citation. This work is evidence alone.

In this work we present a methodology for recovering the degree sequence and the triangle sequence (per edge) under a uniform edge-sampling scenario where for an undirected graph $G$, a sample is constructed by uniformly sampling each edge of $G$ with probability $p$. First, we build on methods by Ganguly et al~\cite{ganguly2017estimation} who recover the degree distribution from node-sampled networks using a Bayesian approach and we extend this to edge-sampled networks. We address the problem of finding an appropriate prior degree distribution by proposing two different ways to construct a prior. We further extend this Bayesian approach to estimating the edge triangle count (the number of triangles associated with each edge) and the total triangle count. 

We find that our Bayesian method outperforms the standard scale-up method at estimating the degree sequence, particularly in small $p$ scenarios where as few as 10\% of the edges remain. Moreover, the priors we use do not make any assumptions about the original degree distribution. For estimating the triangle per link count, in 3 out of the 5 network datasets we use, a Poisson prior achieves similar performance to a correct prior.

This paper is structured as follows. First, in~\cref{sec:properties} we describe the edge sampling procedure and derive properties of graphs that have been sampled in this way. Then in~\cref{sec:estimators} we introduce the various estimators used for these properties, with~\cref{sec:priors} showing how to construct a prior for the Bayes estimators. Finally in~\cref{sec:results} we present our results on recovering these properties on synthetic and real datasets. We discuss the implications in~\cref{sec:conclusion}.

Note that this paper is an extended version of previous work~\cite{arnold2023reconstructing}; in~\cref{sec:properties} we perform a deeper analysis of how the expectation and variance of quantities in the sampled graph relates to the structure of the original graph and how this may affect estimate quality, in~\cref{sec:results} we investigate the empirical bias of our estimators with respect to the different datasets, and we include results for a significantly larger dataset.

% {\color{blue}
% {\bf in case we need to refer to the total number of nodes}
% methods to evaluate the number of missing nodes~\cite{katzir2011estimating,Jianguo12,bianconi2022grand}
% }
\section{Related Work}\label{sec:relatedwork}Sampling of complex networks in general is a well studied problem. One point of interest is how well sampling preserves different properties, such as node rankings in Twitter networks~\cite{morstatter2013sample}, temporal features~\cite{ahmed2013network} and scaling properties~\cite{leskovec2006sampling}. These works have aimed also at designing sampling schemes specifically to preserve a given quantity. Other works have used sampling to estimate quantities on graphs that are prohibitively large to work with in their entirety, with a focus on triangle counting~\cite{tsourakakis2009doulion,antunes2021sampling,stefani2017triest} or other motifs~\cite{klusowski2018counting,bhattacharya2022motif}.

Two recent works studied the problem of recovering a network's degree distribution working from a small sample, first posed by Frank~\cite{frank1971statistical} in his PhD thesis in 1971. The first by Zhang et al~\cite{zhang2015estimating} frames it as an inverse problem involving the vector of observed degree counts and a linear operator representing the sampling scheme. The second by Ganguly et al~\cite{ganguly2017estimation} uses a range of estimators for individual vertex degrees in node-sampled networks; simple scale-up estimators, risk minimisation estimators and Bayes posterior estimates. Antunes et al~\cite{antunes2021sampling} whose work was on sampling methods for estimating the triangle distribution, studied the $n=1$ sample size problem as restricted access scenario as a case study. Other than this, little attention has been given specifically to these restricted access problems, noted in~\cite{zhang2015estimating}. 

The problem of reconstructing triangles from sampled networks has also received attention. Tsourakakis et al~\cite{tsourakakis2009doulion} were motivated by the speed-up that can be obtained by working on sampled not complete data. Their method counts the number of triangles in the graph $G'$ sampled from a full graph $G$ retaining edges independently with probability $p$ (the setting for this paper). They proved that simply multiplying by $1/p^3$ on the observed triangles in the sample $G'$ gives an unbiased estimator for the original number of triangles in $G$. Lim et al~\cite{lim2018memory} developed a streaming estimator that can work in a single pass over data using a similar algorithm. The extend~\cite{tsourakakis2009doulion} to graphs that allow multiple edges between the same node pair. 

A related problem is reconstructing network structure from unreliable or noisy data, such as social networks constructed from reported friendships, which are well known for having missing or spurious edges due to the different interpretations of ``friendship"~\cite{young2020bayesian}. Young et al~\cite{young2020bayesian} address this using a Bayesian approach for finding posterior probabilities for an edge's existence given the measurements obtained. Newman~\cite{newman2018network} use a Bayesian approach involving the empirical false and true positive rates of observing an edge from the data.

While it does not deal directly with reconstruction, \cite{bianconi2005loop} gives an estimate of the number of triangles in a network as
$$\hat{T} = \frac{1}{6}\left[\frac{\sum_k (k-1)k P(k)}{\overline{k}}\right]^3,$$
where $\hat{T}$ is the estimated number of triangles, $P(k)$ is the number of nodes of degree $k$ and $\overline{k}$ is the mean degree. This would allow an estimate of the number of triangles to be derived if the degree distribution is known (or can be estimated). However, the estimate is only reasonable if nodes are connected independently, that is the probability that node $i$ connects to node $j$ is proportional to $k_ik_j$ their degrees. In~\cite{bianconi2005loop} it is shown that this is a reasonable approximation for the Internet Autonomous Systems network but it is not a useful approximation for the majority of networks. 
\section{Properties of edge sampled graphs}\label{sec:properties}Let $G=(V,E)$ be an undirected simple graph with vertex set $V = \{v_1, \dots , v_N\}$ and edge set $E = \{e_1, \dots, e_M\}$. Consider a sampling regime where each edge $e_l \in E$ is included in the sampled graph with probability $p \in [0,1]$, and each vertex $v_i \in V$ is included if any edge incident to it is included. Denote this sampled graph $G'=(V', E')$, where $V' \subseteq V$ and $E' \subseteq E$. Let the sizes of $V'$ and $E'$ be $N'$ and $M'$ respectively. This is known as \emph{incident subgraph sampling}~\cite{leskovec2006sampling}. 
\subsection{Degree}~\label{sec:deriveddegree}
Let $k_i$, respectively $k_i'$ denote the degree of a node $v_i \in G$ and $G'$ respectively. Then $k_i'$ follows a binomial distribution $k_i' \sim B(k_i, p)$, with conditional probability given by
\begin{equation}
    \mathbb{P}(k_i' = k' | k_i = k) = \binom{k}{k'} p^{k'} (1-p)^{k-k'},\label{eqn:degreebinomial}
\end{equation}
with expectation 
$\mathbb{E}(k_i' | k_i =k) = kp$
and variance $\Var(k_i'|k_i=k) = kp(1-p)$.
%using the standard formulae for the expectation and variance of a binomial random variable, respectively.
The probability that node $v_i \in V$ of degree $k_i$ has degree 0 in $G'$ is given by
$\mathbb{P}(k_i' = 0) = (1-p)^{k_i}$.

Nodes in $G$ that become isolated as part of the sampling process are invisible to observers of $G'$ and should be considered removed. In this way, let $\delta_i$ be the indicator random variable representing the removal of node $v_i$ from $G$, with probability $(1-p)^{k_i}$.
\begin{figure}[htbp]
    \centering
    \includegraphics[width=0.5\linewidth]{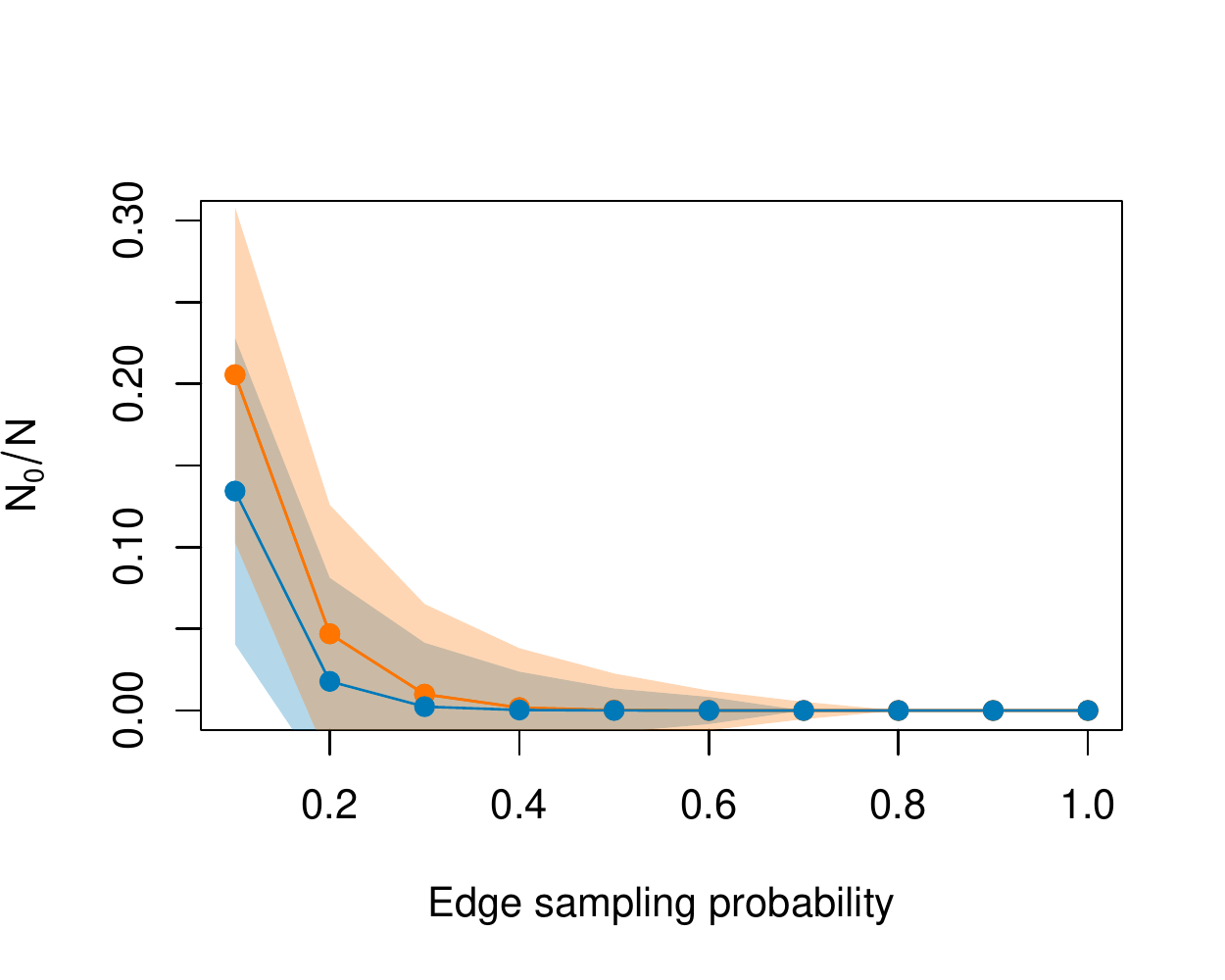}
\caption[Removed nodes in ER and BA networks by edge-sampling]{Proportion of nodes removed in Erd\H{o}s-R\'enyi and Barab\'asi-Albert graphs, of size 1000 nodes and 10000 (ER), 9900 (BA) edges (see~\cref{tab:datasets}), using the edge-sampling procedure. Shown is the mean value $N'_0/N$ for 15000 sampled graphs for each value of $p$ with standard deviation error bars (one above and one below). The y-axis is slightly truncated since the ratio is a positive number.}
    \label{fig:removednodes}
\end{figure}
Then, the expected number $N'_0$ of removed nodes from $G$ is given by
\begin{align}
   \E(N_0) &= \sum_{i=1}^N \E(\delta_i) \notag \\
   &= \sum_{i=0}^N (1-p)^{k_i} \notag \\
   &= \sum_{k\geq 1} (1-p)^{k}N_k \label{eqn:deg0exp}
\end{align}
% $\E(N'_0) = \sum_{i=1}^N \E(\delta_i) = \sum_{k\geq 1} (1-p)^{k}N_k$,
where $N_k$ is the number of vertices in $G$ of degree $k$. This is dependent on the degree distribution; the leading constant term for small $p$ is $N_1$, the number of degree $1$ nodes in $G$. 
This brings to mind the friendship paradox~\cite{feld1991your}, where a node incident to a randomly chosen edge will on average have a higher degree than a randomly chosen node.
From figure~\ref{fig:removednodes} we see 
Barab\'asi-Albert~\cite{barabasi1999emergence} networks experience more node removal than Erd\H{o}s-R\'enyi~\cite{erdHos1960evolution} networks.

\Cref{eqn:deg0exp} is a special case of the expression Dubois et al~\cite{dubois2012effect} derived for the number of nodes of degree $k'$ in an edge-sampled graph $\E(N'_{k'}) =  \sum_k^N N_k \binom{k}{k'} (1-p)^{k-k'}p^{k'}$ with $k'=0$. They note that the number of nodes of degree $k'$ is the sum of variables that are \emph{not} independent of each other and that the resulting degree of a node after edge sampling is intrinsically linked with the resulting degree of its neighbours.

Indeed, we find that the variance in the number of nodes removed $N'_0$ is given by
\begin{align*} 
    \Var(N'_0) &= \Var\left(\sum_{i=1}^N \delta_i \right) = \sum_{i=1}^N \Var(\delta_i) + \sum_{1\leq i\neq j \leq N} \Cov(\delta_i, \delta_j) \\
        &=\sum_{i=1}^N (1-p)^{k_i} \left[1-(1-p)^{k_i}\right] \notag\\
        &\qquad+ \sum_{1 \leq i \neq j\leq N}  (1-p)^{k_i + k_j - A_{ij}} - (1-p)^{k_i + k_j}. \\
%        &= \sum_{k\geq 0} (1-p)^k\left[1 - (1-p)^k\right]N_k \notag \\
%        &\qquad + \sum_{ k, k' \geq 0} \left[(1-p^{k + k'-1} - (1-p)^{k + k'}\right] N_{k,k'}
        &= \sum_{k\geq 0} N_k\left[ (1-p)^k - (1-p)^{2k} \right] + \sum_{k,k'\geq 0} N_{k,k'} p(1-p)^{k + k' - 1}
\end{align*}
where $N_{k,k'}$ is the number of edges connecting vertices of degree $k$ and $k'$. Treating $p$ as a small constant, the dominating terms are $N_1$ and $N_{1,1}$, meaning that the variance is dependent on both the number of low degree nodes and degree-degree correlations among low degree nodes. 
\begin{figure}
    \centering
    \includegraphics[width=0.5\linewidth]{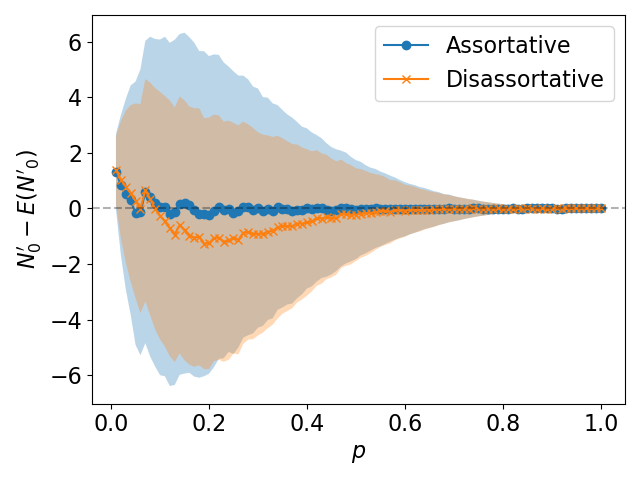}
    \caption{Difference between the expected (using equation~\ref{eqn:deg0exp}) and true number of nodes removed in the edge sampling process for two networks of the same power-law degree distribution but different degree assortativities. The mean is shown over 1000 sampled graphs for each value of $p$ with standard deviation error bars (one deviation above and one below.) }
    \label{fig:zerosAssortative}
\end{figure}
In figure~\ref{fig:zerosAssortative}, we study two networks with the same power-law degree distribution but rewired in two different ways using~\cite{zhou2007structural} to obtain a maximally assortative and maximally disassortative version of that network. Plotted is the difference between number of nodes removed and the expected number of nodes removed using equation~\ref{eqn:deg0exp}, finding a slight small empirical bias for the disassortative network. With all these things in mind, estimating the number of nodes removed by an edge sampling process can therefore be expected to be nontrivial and we do not address it in this work.
% {\color{blue}
% Dubois et al~\cite{dubois2012effect} noticed that the average degree $E(k’)=\sum_k^N k' {k\choose k’} (1-p)^{k’-k}p^{k'}$ of a node after edge removals, its final degree is connected with the final degree of its neighbours hence the sum of independent variables gives a good approximation to the average degree, but is not exact. This can be observed when  the probability of removal is large and evaluating the number of isolated nodes.  {\bf we can add a subgraph for fig 1}

% It is expected that reconstructing the node degree when many links are missing it would have some inaccuracies as the probabilistic model is not completely accurate.
% }
\subsection{Triangles}\label{sec:triangles}
Let $T_l$ be the number of triangles in $G$ which include edge $e_l \in E$. Then the number of triangles in $G$, denoted by $T$, is given by
\begin{equation}\label{eqn:totaltri}
    T = \frac{1}{3} \sum_{e_l \in E} T_l
\end{equation}
where the factor of $\frac{1}{3}$ is present because each triangle in the sum is counted three times, once for each link.

Let $T_l'$ be the number of triangles which include edge $e_l$ in the sampled graph $G'$, defining $T_l' = 0$ if $e_l \notin E'$. In the case that edge $e_l$ remains in the sampled network, then each triangle that includes $e_l$ will remain in the sampled network if and only if the other two edges remain; this occurs with probability $p^2$. There are $T_l$ such triangles, so the number of these which remain in the sampled network is binomially distributed with $T_l$ trials and probability $p^2$. That is,
\begin{equation}\label{eqn:edgein}
    \mathbb{P}(T_l' = t' | T_l = t, e_l \in E') = \binom{t}{t'} p^{2t'} (1-p^2)^{t-t'}.
\end{equation}
In the case that $e_l$ does not remain in the sampled network, the following holds:
\begin{equation}\label{eqn:edgeout}
    \mathbb{P}(T_l' = t' | T_l = t, e_l \notin E') = \delta_{0,t'}
\end{equation}
where $\delta_{0,t'}$ is the Kronecker delta function, taking the value of 1 if $t'=0$ and 0 otherwise (since we defined $T_l'=0$ if $e_l \notin E'$). 

% The law of total probability states that if events $B_1, B_2, \dots, B_n$ partition a sample space $S$, then for any event $A \subset S$,
% \begin{equation}\label{eqn:lotp}
%     P(A) = \sum_{j=1}^n P(A|B_j)P(B_j).
% \end{equation}
We can use the law of total probability to remove the conditioning on $e_l$ from \cref{eqn:edgein,eqn:edgeout} and find $\mathbb{P}(T_l' = t')$, the probability mass function for the number of triangles in the sampled graph link $e_l$ participates in.
\begin{align}
    \mathbb{P}(T_l' = t' | T_l = t) &= \mathbb{P}(T_l' = t' | T_l = t, e_l \in E') \mathbb{P}(e_l \in E') \notag\\
    &+ \mathbb{P}(T_l' = t' | T_l = t, e_l \notin E') \mathbb{P}(e_l \notin E') \notag \\
    &= p\binom{t}{t'} p^{2t'} (1-p^2)^{t-t'} + \delta_{0,t'} (1-p).\label{eqn:T'givenT}
\end{align}
%letting $B_1$ and $B_2$ in \cref{eqn:lotp} be the events that $e_l \in E'$ and $e_l \notin E'$, with probabilities $p$ and $(1-p)$ respectively, and $A$ being the event $T_l' = t'$ given that $T_l = t$.
Therefore, the conditional probability mass function for $T_l'$ given $T_l$ is given by~\cref{eqn:T'givenT}.

The expected value of $T_l'$ given $T_l$ is given by
\begin{align}
    \mathbb{E}(T_l'|T_l=t) &= \sum_{t'=0}^{t} t'\mathbb{P}(T_l'=t'|T_l = t) %\label{eqn:expectedval} 
    \notag \\
    &= \sum_{t'=0}^t t' \left[p\binom{t}{t'} p^{2t'} (1-p^2)^{t-t'} + \delta_{0,t'} (1-p)\right] \notag \\
    %\label{eqn:expectedtrimessy} \\
    %\mathbb{E}(T_l'|T_l=t)
   % &= \sum_{t'=0}^t t' \left[p\binom{t}{t'} p^{2t'} (1-p^2)^{t-t'} \right] \\
    &= p \sum_{t'=0}^t t' \binom{t}{t'} p^{2t'} (1-p^2)^{t-t'} 
 = p\times p^2t =p^3t \label{eqn:aha} 
\end{align}
where \cref{eqn:aha} comes from noting that the sum in the left hand side precisely evaluates the expected value of a binomial random variable with $t$ trials and probability $p^2$.

Let $T'$ be a random variable representing the triangle count of $G'$, then

\begin{align}
    \mathbb{E}(T') = \mathbb{E}\left[\frac{1}{3} \sum_{e_l \in E} T_l'\right] = \frac{1}{3} \sum_{e_l \in E} p^3 T_l = p^3 T \label{eqn:expectedtri}
\end{align}
where $T'$ is the triangle count of the sampled network $G'$.

\begin{figure}[htbp]
    \centering    \includegraphics[width=0.6\linewidth]{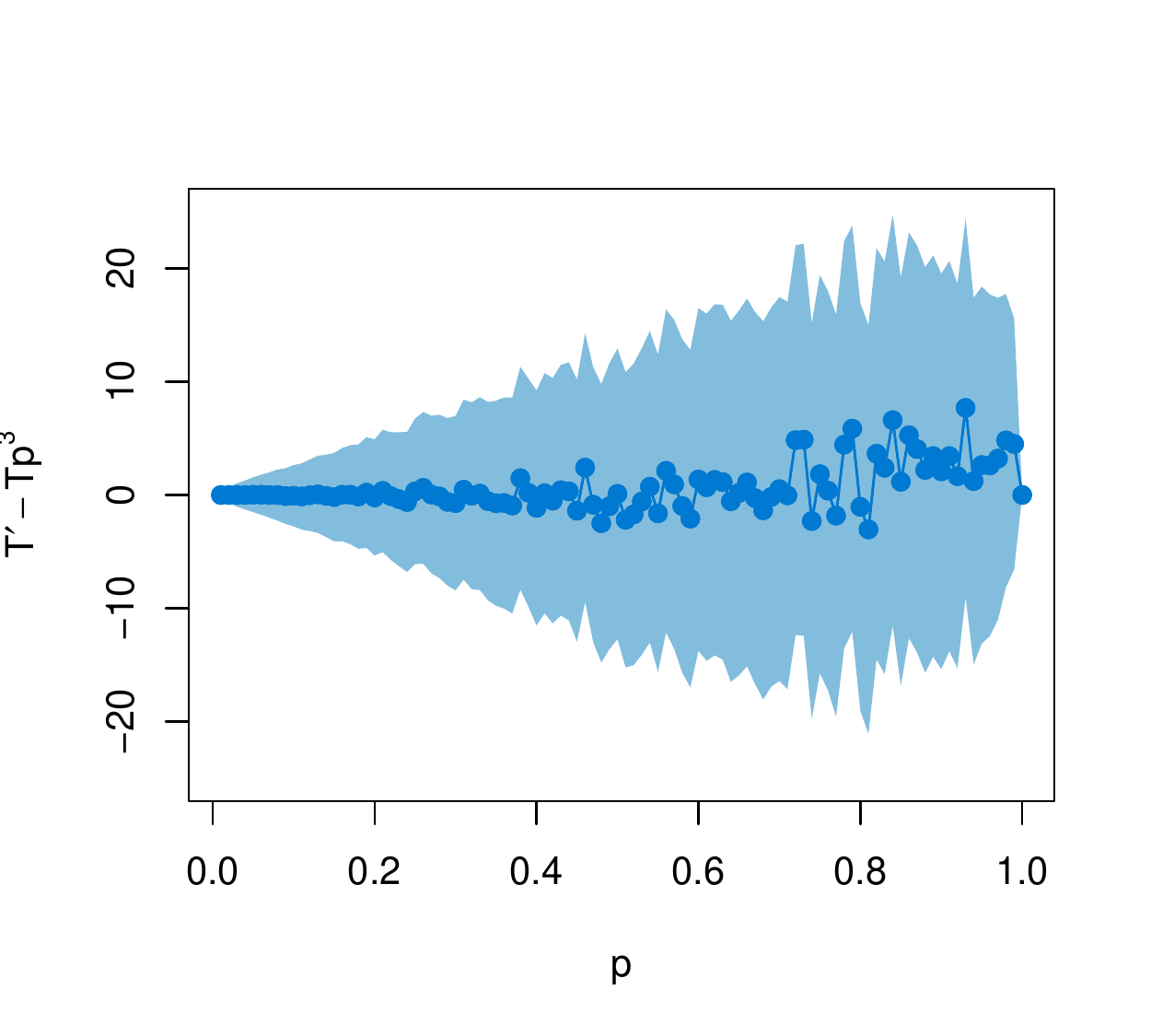}
    \caption[]{Difference between the observed number of triangles and the expected number of trinagles using $p^3T$, on a power law network with $N=1,000$ and $M=5,000$ and degree assortativity $\rho=0.35$. Shown is the mean over 500 sampled graphs for each value of $p$ and with standard deviation error bars (one deviation above and one below.)}
    \label{fig:UglyGraph}
\end{figure}

Similar to the discussion in the previous section, the sum in~\cref{eqn:expectedtri} is not the sum of independent random variables. The quantities $T'_l$ and $T'_m$ for $l\neq m$ are only independent if edges $e_l, e_m \in E$ are not part of triangles which overlap in any way; this can happen not only if they are adjacent edges but if they have triangles dependent on a shared edge. We can therefore expect total triangle count estimates to be more variable in networks with numerous instances of multiple triangles sharing a link; in these cases, the removal of a single link may remove a large number of triangles. This nature of clustered triangle distribution would be likely in a network that  was highly transitive.
As an example, figure~\ref{fig:UglyGraph} shows the average number of triangles $T'$ in a sampled graph minus the estimated number of triangles $Tp^3$.
The network is a power law network with positive transitivity (global transitivity  $\bar C=0.14$ and assortativity $\rho=0.35$). The total number of triangles in the original network is 13,305 and the maximum observed effect is an underestimate of around 8 triangles. This is consistent with low or zero bias in the estimator. Note that the large error bars for large $p$, while may seem counterintuitive, are due to larger variance in $T'$ for these values of $p$ and hence larger standard errors.

The variance of $T_l'$ given $T_l$ is given by
\begin{align}
    \Var(T_l'|T_l=t) &= \mathbb{E}(T_l'^2 |T_l=t) - \mathbb{E}(T_l'|T_l=t)^2 \notag \\
    &= p\sum_{t'=0}^t t'^2\left[ \binom{t}{t'} p^{2t'}(1-p^2)^{(t-t')} \right] - p^6t^2 \notag \\
    &= p\left(\sum_{t'=0}^t t'^2\left[ \binom{t}{t'} p^{2t'}(1-p^2)^{(t-t')} \right] - p^4t^2\right) \notag \\ &\qquad+ p^5 t^2 - p^6t^2 \label{eqn:binomialtrick}\\
    &= p^3t(1-p^2) + p^5t^2 - p^6t^2 \notag \\
    &= p^3t(1 - p^2 + p^2t - p^3t) \notag
\end{align}
where the bracketed term in line~\ref{eqn:binomialtrick} is the variance of a binomial variable with $t$ trials and probability $p^2$.

An argument involving computation of the covariances $\Cov(T_j, T_l)$ shows that the variance of the expected total triangle count of $G'$ given the individual triangle counts $T_1, \dots , T_M$ is given by
\begin{align}
    \Var(T'|T_1, \dots , T_M) &= \frac{1}{9}\bigg[ 3p^3(1-p^2)T + (p^3-p^2)\sum_{e_l \in E} T_l^2  \notag\\ 
    &\qquad+ 6T(p^3-p^6) + 8k(p^5-p^6) \bigg].
    \label{eqn:tdash_var}
\end{align}
where $k$ is the number of triangles which share a link. The full derivation of~\cref{eqn:tdash_var} can be found in the first author's thesis~\cite{arnold2021studying} which also contains derivations for wedge\footnote{A two hop path, a triangle is composed of three wedges.} counts and clustering coefficient.
An expression for this variance conditioned on total triangle count $T$ not edge triangle counts is given in~\cite{tsourakakis2009doulion}.

The number of triangles per node $T_i$ can be obtained from $T_{e_\ell}$ from
$2T_i=\sum_k T_{e_{\ell}=(i,k)\in E}$
meaning the estimators of the edge-sampled network can be extended to evaluate vertex statistics eg local transitivity of the nodes $c_i= \sum_k T_{e_{\ell}=(i,k)\in E}/(k_i(k_i-1))$. 

\section{Estimators for the degree sequence and triangle count}\label{sec:estimators}The previous showed how the distribution of a quantity $X'$ in a sampled graph $G'$ could be calculated as a conditional probability $P(X'=x'|X=x)$ given the unsampled measurement $X=x$. This section aims to estimate the true network quantity $X$ given its sampled counterpart $X'$.
\subsection{Method of moments estimators}~\label{sec:MME}
Let $X$ be a random variable associated with a statistic of $G$ and let $X'$ be that statistic on $G'$ with expected value $\mathbb{E}(X') = f(X,p)$. A naive `scale-up' estimator for $X$ given observed value $x'$ for $X'$ is the solution $\hat{x}$ to the equation $x' = f(\hat{x},p)$, provided a solution exists. Borrowing the terminology from~\cite{ganguly2017estimation}, we will refer to these estimators as \emph{method of moments estimators} (MME).
\subsubsection{Degree}
For a node of degree $k_i'$ in $G'$, the MME for $k_i$ is given by $k_i'/p$. This is an unbiased estimator with mean $\E(k_i'/p) = \frac{1}{p} k_i p = k_i$ and variance $\Var(k_i'/p) = \frac{1}{p}k(1-p)$. Nodes with the lowest possible degree (one) in the sampled graph are estimated as having degree $1/p$ in the unsampled graph so as $p$ decreases, the estimation of low-degree nodes becomes poorer. 
\subsubsection{Triangle count}
The expected triangle count $\mathbb{E}(T_l')$ of edge $e_l$ is $p^3T_l$. If in addition, $e_l$ remains in $G'$, its expected triangle count is given by $p^2 T_l$. Therefore the MME for $T_l$ is $p^{-3}T_l'$ or $p^{-2}T_l$, without and with the conditioning respectively. Similar to the MME for degree, it provides poor estimates for edges that have a low triangle count, as it disallows any estimates of $T_l$ in the range $(0, 1/p^2)$.

Similarly, an MME estimate proposed by Tsourakakis et al~\cite{tsourakakis2009doulion} for the total triangle count of a network is $p^{-3} T'$, which has expected value $\E(p^{-3} T') = T$. They found that this estimator has variance $\frac{1}{p^6}\left((p^3-p^6)T + 2k(p^5 - p^6)\right)$, where $k$ is the number of pairs of triangles which share a link.

\subsection{Bayes estimator}~\label{sec:bayesestimators}
\noindent This estimator relies on  Bayes theorem, giving
\begin{equation*}
    \mathbb{P}(X=x|X'=x') = \frac{\mathbb{P}(X'=x'|X=x) \mathbb{P}(X=x)}{\mathbb{P}(X'=x')}.
\end{equation*}
$\mathbb{P}(X'=x'|X=x)$ is the \emph{likelihood} which is determined by the edge sampling procedure and is known. $P(X=x)$ is the \emph{prior} function which will be denoted by $\pi(x)$; this is in general not known.

A posterior estimate for $X$ given $X'$ can then be given as the expected value
\begin{equation*}
    \E(X|X'=x') = \frac{\sum_x x\mathbb{P}(X'=x'|X=x)\pi(x)}{\mathbb{P}(X'=x')}.
\end{equation*}
The immediate question arises of how to deal with the prior $\pi(x)$, as this may involve making assumptions about the structure of $G$. This will be discussed case by case for the degree and triangle count.
\subsubsection{Degree}\label{sec:degreebayes}
Using the likelihood function for the degree from \cref{eqn:degreebinomial}, a posterior estimate for the degree of node $v_i$ given it has degree $k_i'$ in $G'$ is
\begin{equation} %\label{eqn:bayesdegree}
    \mathbb{E}(k_i|k_i'=k') = \frac{\sum_{k=k'}^{\infty} k \binom{k}{k'}(1-p)^{k} \pi(k) }{\sum_{k=k'}^{\infty} \binom{k}{k'}(1-p)^{k} \pi(k)}
\end{equation}
where $\pi(k)$ is a prior for the degree distribution $P(k)$ of $G$.

\subsubsection{Triangle count}
Using the likelihood function from \cref{eqn:edgein}, a posterior estimate for the triangle count of edge $e_l$ in $G$ given it remains in $G'$ is 
\begin{align}\label{eqn:posteriortriangles}
    \mathbb{E}(T_l|T_l'=t',e_l \in G') 
    &= \cfrac{\sum_{t=t'}^{\infty} t \binom{t}{t'}(1-p^2)^{t} \pi(t) }{\sum_{t=0}^{\infty} \binom{t}{t'}(1-p^2)^{t} \pi(t)  }
\end{align}
%&= \cfrac{\sum_{t=t'}^{\infty} t \binom{t}{t'}p^{2t'}(1-p^2)^{t-t'} \pi(t) }{P(t')} \notag \\
    % &= \cfrac{\sum_{t=t'}^{\infty} t \binom{t}{t'}p^{2t'}(1-p^2)^{t-t'} \pi(t) }{\sum_{t=0}^{\infty} \binom{t}{t'}p^{2t'}(1-p^2)^{t-t'} \pi(t)  } \notag \\
where $\pi(t)$ is a prior for the proportion of edges with triangle count $t$.~\footnote{In experimental runs, the native binomial functions introduced numerical inaccuracies for large powers of $(1-p)$. Therefore, an equivalent evaluation of binomial probabilities using the log-gamma function and laws of logs was used in practice.}

To establish the total triangle count, summing the value of this estimator over the remaining edges in $G'$ and dividing by 3, as in \cref{eqn:totaltri}, will provide an underestimate for the total triangle count of $G$, since there are potentially many missing edges in $G'$. To mitigate this, we scale this factor up to the estimated number of edges in $G$. That is, our estimate of the total triangle count becomes
\begin{equation*}
   \hat{T} = \frac{1}{3p} \sum_{e_l \in G'} \hat{T}_l. 
\end{equation*}
\subsection{Remarks on our estimators}
It is worth highlighting that up to now, our degree estimators $\hat{k_i}, \hat{k_j}$ for nodes $i$ and $j$ will take the same value if $k_i'$ and $k_j'$ are the same. That is, only the node's degree and the degree sequence of the sampled network are taken into account and no other aspect of the network's structure. It also means that estimates will be very coarse; if the sampled network has maximum degree $k$ then the estimated degree sequence will take up to $k$ different values. The same is true for the triangle per link estimators.
\section{Constructing a prior}\label{sec:priors}The Bayes estimators for the degree and triangle per link sequences require the choice of a prior $\pi(k)$ and $\pi(t)$ for the degree distribution and triangles per link distribution respectively. In this section, we propose methods for constructing priors.
\subsection{Degree distribution}
A prior could be obtained from chosen family of distributions such as the Zipf distribution or a power law distribution, but this baked-in assumption may not be desirable. Furthermore, it has been shown that the distribution of a sampled network may not even follow the distribution of the true network~\cite{stumpf2005subnets}. Therefore, we propose two different methods of constructing a prior which do not make assumptions about the degree distribution of the true network.

\subsubsection{Monte Carlo minimisation method}
First, it is possible to estimate the prior using a Monte Carlo method to minimise the $\ell_2$ norm of the error, which in this work we will refer to as the \emph{minimisation method}. In this approach, we find a degree sequence $\{\kappa_i\}$ which minimises
$\min\left( \sum_i||p\kappa_i - k'_i||_2^2\right)$. This is done with the restrictions that the degree $\kappa_i$ is an integer number with $\kappa_i \geq k_i'$ and
\begin{equation}
    \sum_i \kappa_i = \lfloor 2M'/p \rfloor.\label{eqn:kappaequality}
\end{equation} To do this, we start with $\kappa_i = \lfloor k_i'/p \rfloor$. If the sum $\sum_{v_i \in V'} \kappa_i$ of the estimated degrees is not equal to the estimated number of links $ \lfloor 2M'/p \rfloor$ then we increment or decrement the degree of nodes chosen uniformly at random until equality holds. Then we construct a graph with degree distribution given by $\{\kappa_i\}$ and rewire its links at random by a single edge swap (incrementing the degree of one node chosen at random and decrementing the degree of another node chosen at random) for a large number of iterations (15000 in our case), accepting each proposed rewire if it decreases the $\ell_2$ error. Note that if the initial estimate $\kappa_i$ satisfies equation~\ref{eqn:kappaequality}, then this is a global minimum for the $\ell_2$ norm. In particular, this happens when $k_i'/p$ is an integer for all $i$.

\subsubsection{Link cascade method}
The MME (and hence sometimes the minimisation method) cannot estimate the degree $k_i\approx k'_i/p$ when $k'_i=0$; that is, the lowest possible estimated degree is $1/p$. If a good estimate for the original number of nodes is known then another prior for capturing these low degree nodes is constructed by ``cascading" links from high degree to low degree nodes. 
This is a heuristic method based on the observation, that when $k'_i=0$ the estimation is poor, so the idea is to create a prior that has the original number of nodes and there are no zero degree nodes. As with the  minimisation method, we start with an estimated degree sequence $\kappa_i = \lfloor k_i'/p \rfloor$, redistributing links as before if the total estimated degree does not match twice the estimated number of links. Then, we place the nodes in descending order based on their estimated degree, with the knowledge of the original number of nodes in the network being used to append placeholder nodes which would have been removed by the sampling process. Finally, we pick the first occurring node in this list with degree zero, and increment this degree by simultaneously decrementing the degree of the node directly before it, this operation is to conserve the total number of links. This step is performed iteratively until there are no degree zero nodes. This heuristic method is fast as it does not require the construction of a graph and the shuffling of its links. As the method manipulates only the degree sequence the method is attractive when trying to construct a prior for large networks. Finally, as a comparison point representing the best possible result achievable with the Bayes method, we use a true prior which places the probability mass $\pi(k)$ equal to the proportion of nodes of degree $k$ in the original network.
\subsection{Triangle per link distribution}
The two methods for prior construction of the degree distribution do not immediately translate to an analogue for triangles, and little is known about the triangle per edge distribution as a starting place for selecting a prior. As an initial approach therefore, we use a Poisson distribution $\text{Po}(\lambda)$ with $\lambda = 3\hat{T}/M'$, the average number of triangles per link in the MME estimator. As with the degree distribution, we include a result with a true prior $\pi(t)$ equal to the proportion of links with triangle count $t$ in the original network as a comparison point.
\section{Results}\label{sec:results}\begin{figure}[htbp]
    \centering
    \begin{subfigure}[b]{0.47\textwidth}
    \centering
    \includegraphics[width=\textwidth]{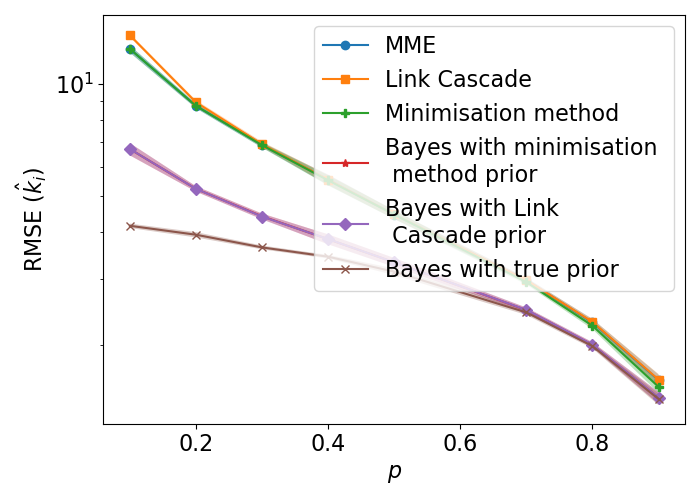}
    \caption{Erd\H{o}s-R\'enyi network}
    \end{subfigure}
    \begin{subfigure}[b]{0.47\textwidth}
      \centering
    \includegraphics[width=\textwidth]{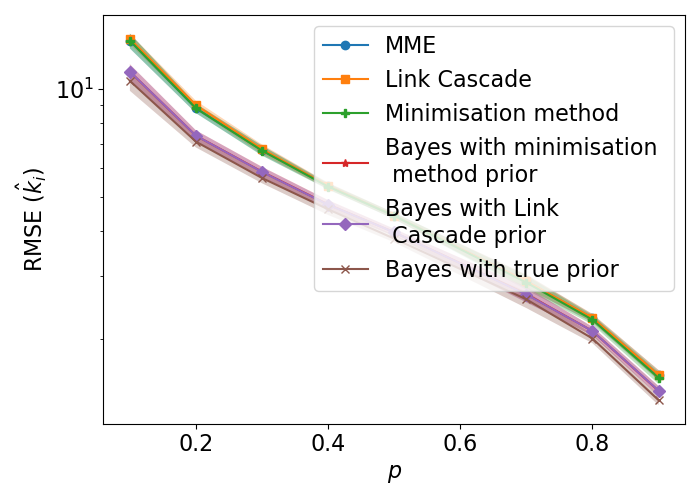}
    \caption{Barab\'asi-Albert network}
    \end{subfigure}  
    \vskip\baselineskip
    \begin{subfigure}[b]{0.47\textwidth}
    \centering
    \includegraphics[width=\textwidth]{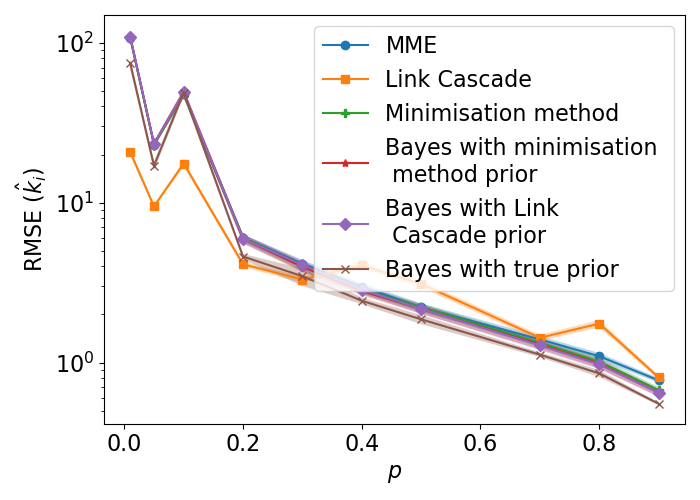}
    \caption{AS Topology network}
    \end{subfigure}
    \begin{subfigure}[b]{0.47\textwidth}
      \centering
    \includegraphics[width=\textwidth]{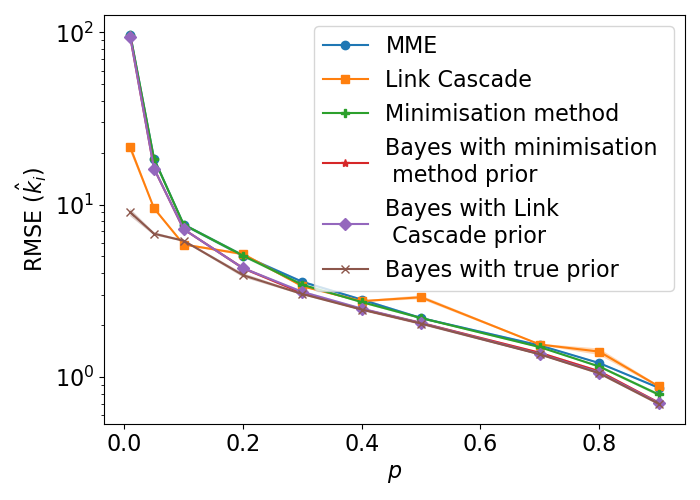}
    \caption{ArXiv Hep-Th collaboration network}
    \end{subfigure}  
    \vskip\baselineskip
    \begin{subfigure}[b]{0.47\textwidth}
    \centering
    \includegraphics[width=\textwidth]{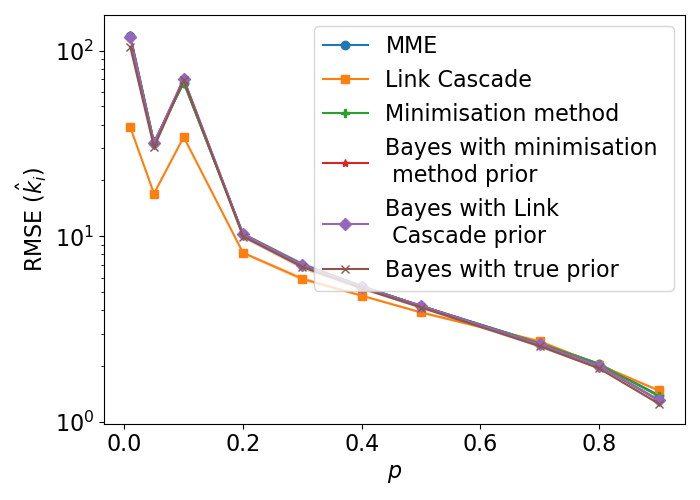}
    \caption{MathOverflow Interactions}
    \end{subfigure}
    \caption{Error in estimation of the true degree sequence. Each value is averaged over 10 experiments with the shaded error bars representing standard deviation and plotted with log-scaled y-axis. The MME overlays the minimisation method (blue, green respectively) in all plots; the same is true of the Bayes methods with minimisation and link cascade priors (red and purple respectively). In (e) all apart from the link cascade method are overlaid. The error bars in some places are very small, in part because we are considering root mean squared error and additionally the y axis is log-scaled.}
    \label{fig:degree_reconstruction}
\end{figure}
To test the capability of our estimators of degree sequence and triangle count, we consider five different starting networks: an Erd\H{o}s--R\'enyi $G(N,M)$ network~\cite{erdHos1960evolution} with $N=1000$ and $M=10000$, a Barab\'asi--Albert network~\cite{barabasi1999emergence} of approximately the same size and density, a real collaboration network from authors who submitted to the ArXiV high-energy theoretical physics category~\cite{newman2001structure} (henceforth Hep-Th for brevity), an Internet autonomous systems topology dataset (henceforth AS)~\cite{chen2002origin} and an interaction network on the MathOverflow sub-forum on StackExchange~\cite{paranjape2017motifs}. A quick reference of some summary statistics can be found in~\cref{tab:datasets}. These datasets were chosen to represent a heterogeneous selection of network types. The ER network has a Poisson degree and edge triangle count distribution and an overall low number of triangles for the network's density. The BA network has a theoretical power law degree distribution and a low triangle count for its density. The Hep-Th and AS networks have a heavy-tailed degree distribution but have very different degree correlations, and the Hep-Th has a higher clustering than the AS network. The MathOverflow dataset is provided as an example of a larger network. For each of these datasets modelled as a graph $G$, we take an edge-sampled network $G'$ with edge sampling probability $p$, for $p = 0.1, 0.2, \ldots , 0.9$ and from this, reconstruct the degree sequences, edge triangle counts and total triangle counts using our estimators. In the larger datasets (Hep-Th, AS topology, StackExchange) we examine also the cases $p=0.01$ and $p=0.05$ to consider the most extreme scenarios.
\begin{table}[htb]
    \centering
    \begin{tabular}{c c c c c c c c}
    \hline
        Dataset & $N$ & $M$ & $k_{\text{max}}$ &$ \rho $ & $T$ & $\Bar{C}$ & $\Bar{T_l}$ \\
        \hline
        Erd\H{o}s-R\'enyi & 1000 & 10000 & 35 & -0.002 & 1373 & 0.021 & 0.41 \\
        Barab\'asi-Albert & 1000 & 9900 & 170 & -0.041 & 6099 & 0.063 & 1.85 \\
        Hep-Th & 5835 & 13815 & 50 & 0.185 & 10624 & 0.506 & 2.31 \\
        AS Topology & 11174 & 23409 & 2389 & -0.195 & 19894 & 0.296 & 2.55 \\
        MathOverflow & 24759 & 187985 & 2172 & -0.215 & 1403896 & 0.313 & 22.4 \\
        \hline
    \end{tabular}
    \caption{Original statistics of network datasets used prior to sampling. Shown is the number of nodes $N$, number of edges $M$, maximum degree $k_{\text{max}}$, degree assortativity $\rho$, number of triangles $T$, average node clustering coefficient $\Bar{C}$~\cite{watts1998collective}, and average number of triangles per link $\Bar{T_l}$.}
    \label{tab:datasets}
\end{table}
In the degree distribution experiment, we reconstruct the degree $k$ of nodes in $V'$ using our chosen estimators $\hat{k}$, and compute the root mean squared error of the degree sequences as $\text{RMSE}(\hat{k}) =\sqrt{\frac{1}{N'} \sum_{i=1}^N (k_i - \hat{k_i})^2}$. These results are shown in figure~\ref{fig:degree_reconstruction}, showing the mean and s.d. error over 10 experiments. In all but the AS topology network and the Math Overflow network, the Bayes estimator with true prior has the lowest error, though this is included only to show the best possible result that could be obtained with the Bayes method since the true prior is unknowable. The link cascade method performs well for many networks in the realistic cases with low sample rates but poorly for the E-R network. Bayes with link cascade prior is a good all round performer, never being a bad performer and second or third best for most networks. This method assumes knowledge of the number of nodes in $G$ (i.e. the number of nodes pruned by the edge-sampling) so performs better at estimating low degree nodes. This is particularly evident in figure~\ref{fig:degree_reconstruction}(c), performing better than the Bayes approach with true prior. The Monte Carlo minimisation method used as an estimator overlays the MME due to the restriction that the degree sequence is an integer (c.f.~\cref{sec:priors}).

\begin{figure}[htbp]
    \centering
    \begin{subfigure}[b]{0.47\textwidth}
    \centering
    \includegraphics[width=\textwidth]{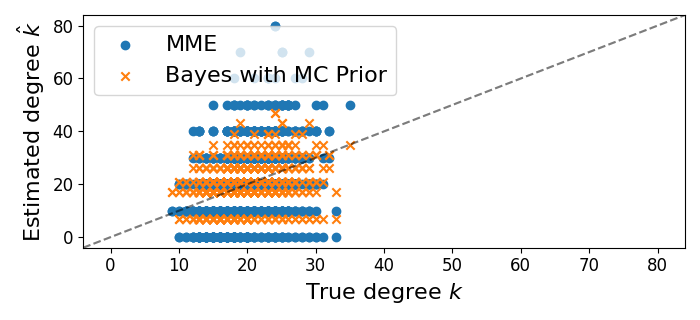}
    \caption{Erd\H{o}s-R\'enyi $p=0.1$}
    \end{subfigure}
    \begin{subfigure}[b]{0.47\textwidth}
    \centering
    \includegraphics[width=\textwidth]{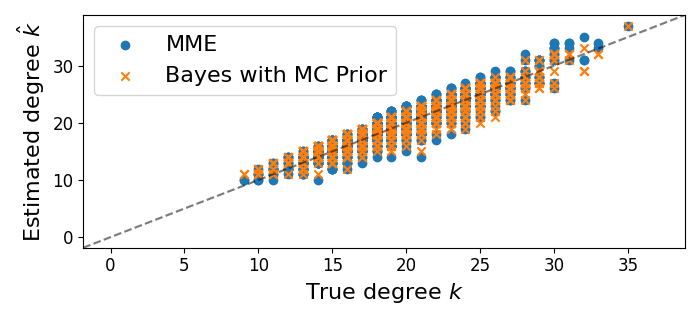}
    \caption{Erd\H{o}s-R\'enyi $p=0.9$}
    \end{subfigure}
    % \vskip\baselineskip
    \begin{subfigure}[b]{0.47\textwidth}
    \centering
    \includegraphics[width=\textwidth]{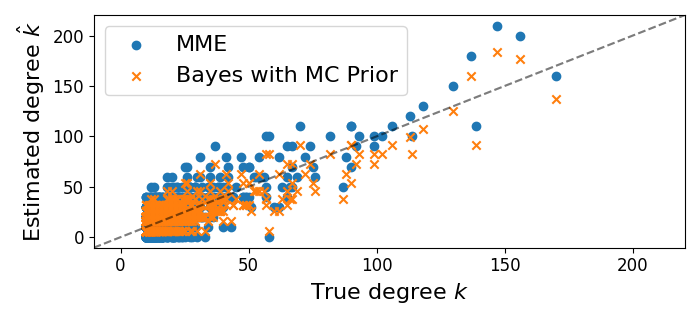}
    \caption{Barab\'asi-Albert $p=0.1$}
    \end{subfigure}
    \begin{subfigure}[b]{0.47\textwidth}
    \centering
    \includegraphics[width=\textwidth]{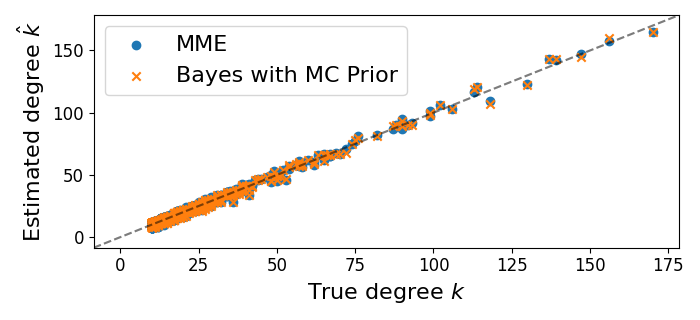}
    \caption{Barab\'asi-Albert $p=0.9$}
    \end{subfigure} 
    % \vskip\baselineskip
    \begin{subfigure}[b]{0.47\textwidth}
    \centering
    \includegraphics[width=\textwidth]{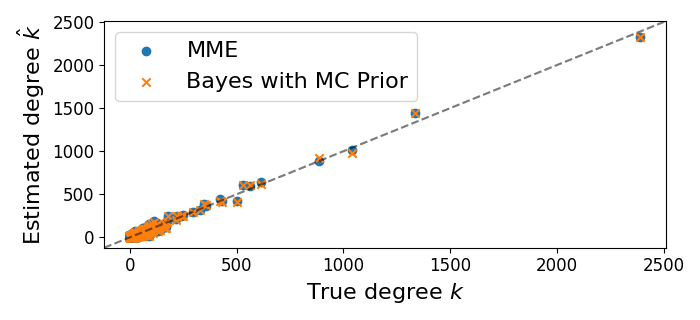}
    \caption{AS topology $p=0.1$}
    \end{subfigure}
    \begin{subfigure}[b]{0.47\textwidth}
    \centering
    \includegraphics[width=\textwidth]{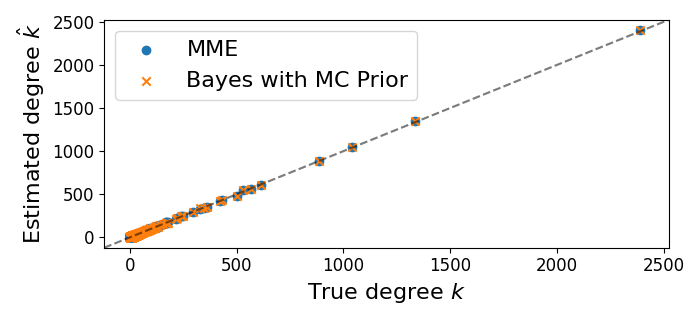}
    \caption{AS topology $p=0.9$}
    \end{subfigure} 
    % \vskip\baselineskip
    \begin{subfigure}[b]{0.47\textwidth}
    \centering
    \includegraphics[width=\textwidth]{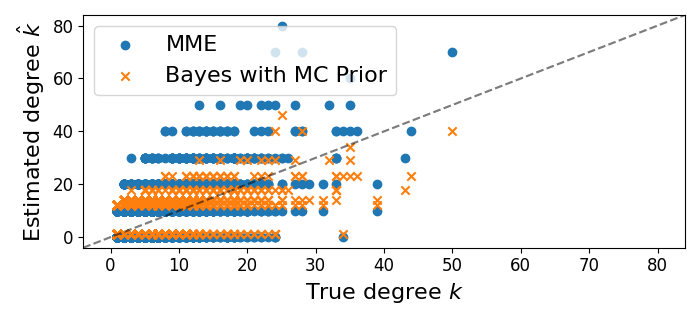}
    \caption{ArXiv Hep-Th $p=0.1$}
    \end{subfigure}
    \begin{subfigure}[b]{0.47\textwidth}
    \centering
    \includegraphics[width=\textwidth]{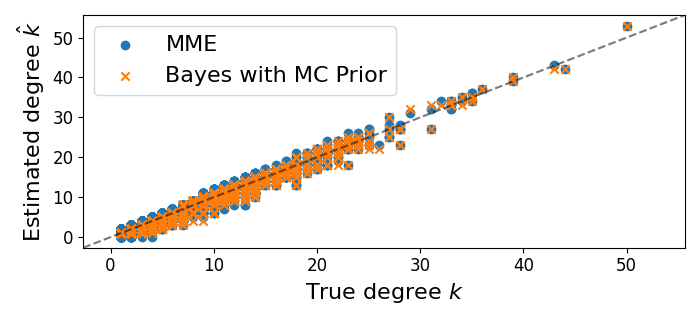}
    \caption{ArXiv Hep-Th $p=0.9$}
    \end{subfigure}
    % \vskip\baselineskip
    \begin{subfigure}[b]{0.47\textwidth}
    \centering
    \includegraphics[width=\textwidth]{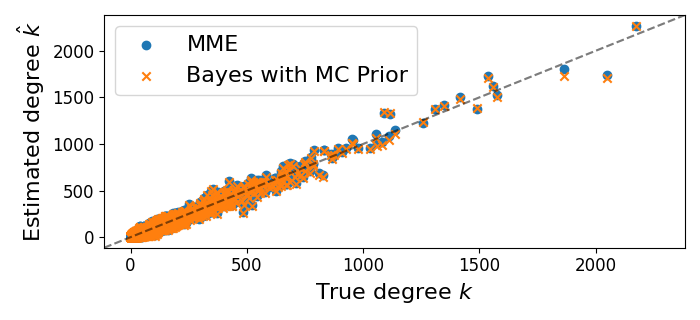}
    \caption{MathOverflow $p=0.1$}
    \end{subfigure}
    \begin{subfigure}[b]{0.47\textwidth}
    \centering
    \includegraphics[width=\textwidth]{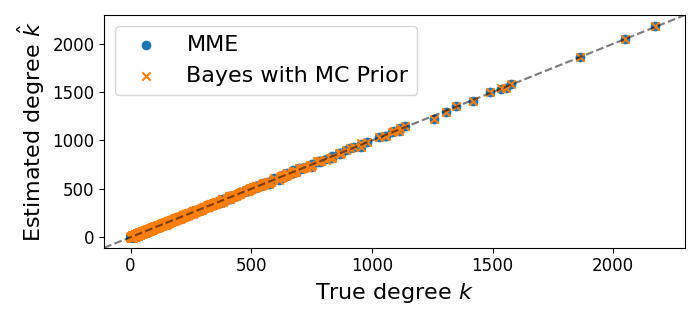}
    \caption{MathOverflow $p=0.9$}
    \end{subfigure}
    \caption{Estimated degree $\hat{k}$ against the true degree $k$ for five different datasets, at a $p=0.1$ sample and $p=0.9$ sample. We compare the method of moments estimator $\hat{k} = k'/p$ with the Bayes posterior estimate with prior obtained using the minimisation method.}
    \label{fig:degree_vs_estimated}
\end{figure}

Figure~\ref{fig:degree_vs_estimated} studies the performance of the Bayes posterior estimate (the one with prior obtained using the minimisation method) against the baseline method of moments estimator in small $p$ and large $p$ scenarios, with the assumption that the number of nodes in the original graph is known. For small values of $p$, our estimates for low degree nodes are very poor. This is due to the information loss; for $p=0.1$ as an example, it is more likely than not for a node of degree 6 or fewer to lose all its edges in the sampling process, and hence the performance is worse in networks with a low maximum degree such as the Erd\H{o}s-R\'enyi and Hep-Ph collaborations graph. The MME is further restricted to taking values that are multiples of 10 for this value of $p=0.1$, so gives very coarse estimates. However, for the $p \leq 0.1$ cases, the Bayes posterior method brings the estimates closer to the correct answer than the MME in all but the AS topology dataset, where the two methods appear to perform similarly.

\begin{figure}[htbp]
    \centering
    \begin{subfigure}[b]{0.46\textwidth}
    \centering
    \includegraphics[width=\textwidth]{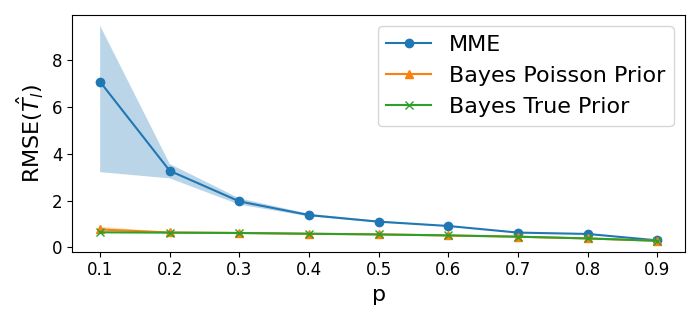}
    \caption{Erd\H{o}s-R\'enyi $T_l$}
    \end{subfigure}
    %\hfill
    \begin{subfigure}[b]{0.45\textwidth}
      \centering
    \includegraphics[width=\textwidth]{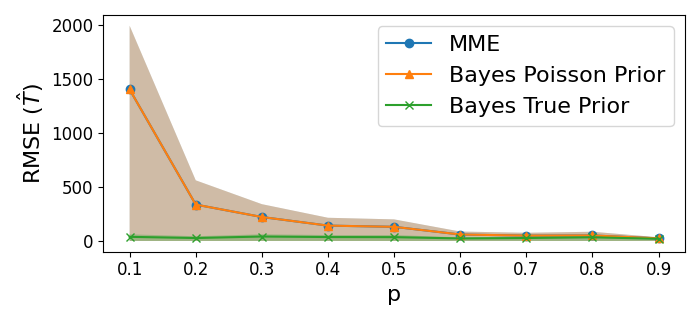}
    \caption{Erd\H{o}s-R\'enyi $T$}
    \end{subfigure}  
    %\vskip\baselineskip
    \begin{subfigure}[b]{0.45\textwidth}
    \centering
    \includegraphics[width=\textwidth]{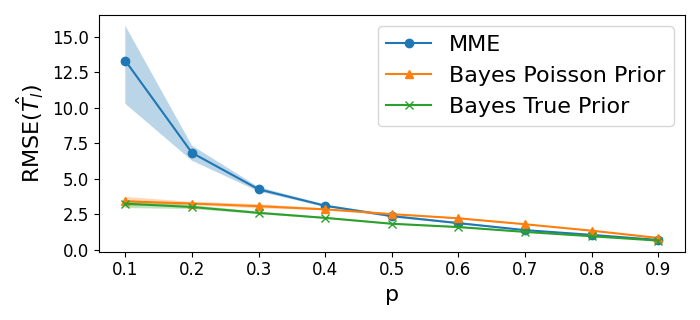}
    \caption{BA $T_l$}
    \end{subfigure}
    %\hfill
    \begin{subfigure}[b]{0.45\textwidth}
      \centering
    \includegraphics[width=\textwidth]{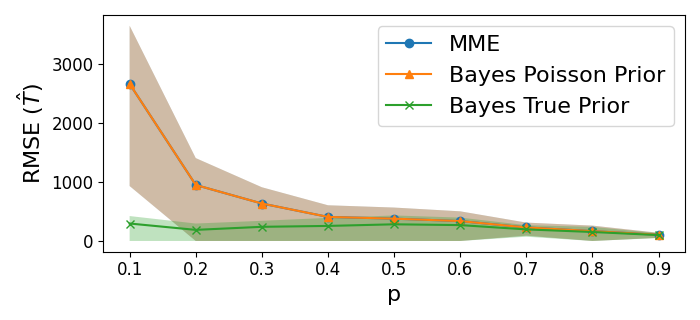}
    \caption{BA $T$}
    \end{subfigure} 
     %\vskip\baselineskip
    \begin{subfigure}[b]{0.45\textwidth}
    \centering
    \includegraphics[width=\textwidth]{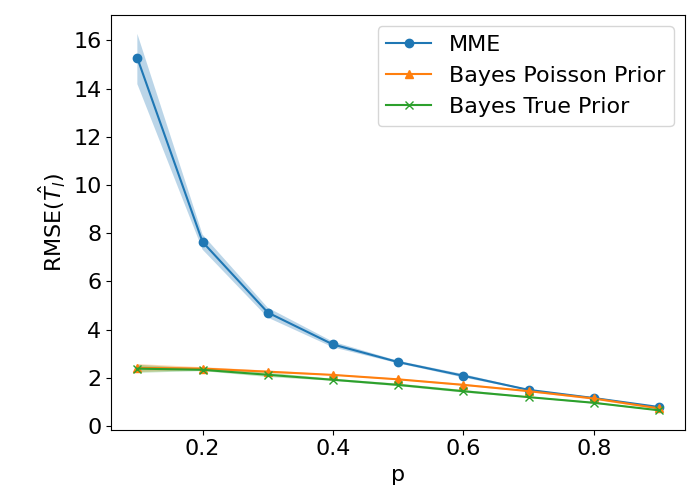}
    \caption{Hep-Th $T_l$}
    \end{subfigure}
    %\hfill
    \begin{subfigure}[b]{0.45\textwidth}
      \centering
    \includegraphics[width=\textwidth]{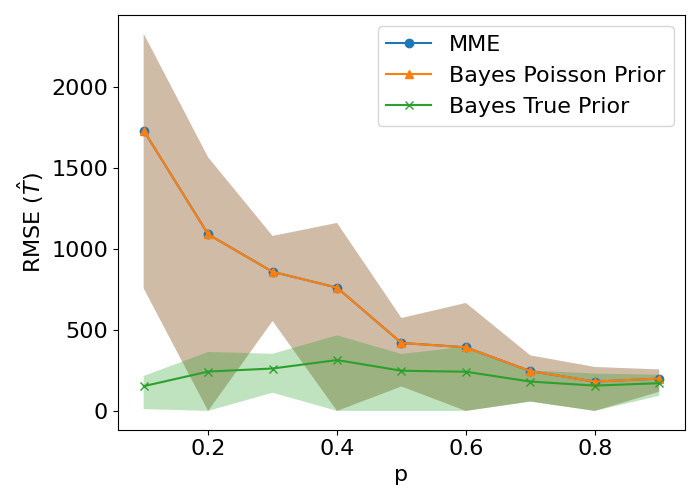}
    \caption{Hep-Th $T$}
    \end{subfigure}  
    %\vskip\baselineskip
    \begin{subfigure}[b]{0.45\textwidth}
    \centering
    \includegraphics[width=\textwidth]{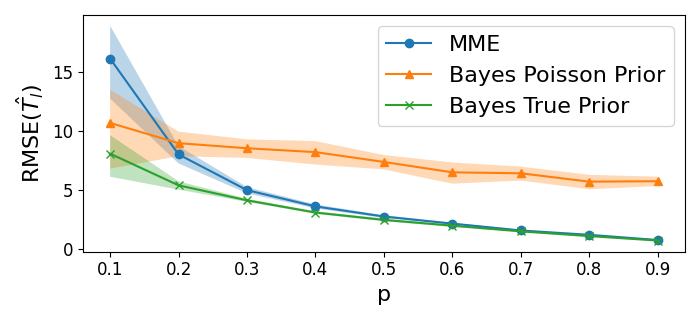}
    \caption{AS Topology $T_l$}
    \end{subfigure}
    %\hfill
    \begin{subfigure}[b]{0.45\textwidth}
      \centering
    \includegraphics[width=\textwidth]{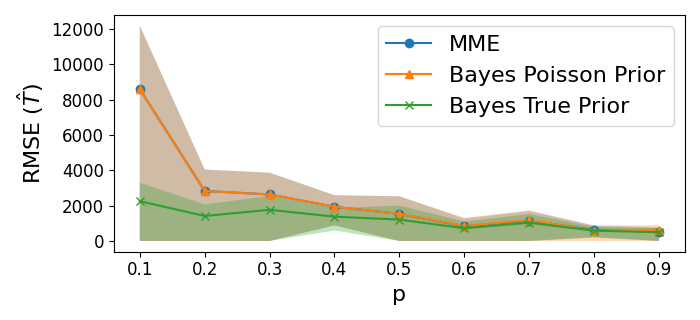}
    \caption{AS Topology $T$}
    \end{subfigure}  
    % \vskip\baselineskip
    \begin{subfigure}[b]{0.45\textwidth}
    \centering
    \includegraphics[width=\textwidth]{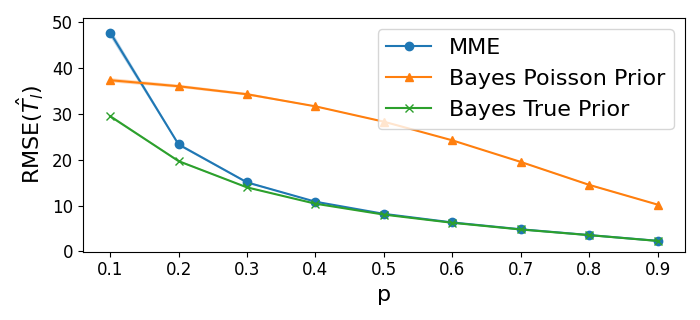}
    \caption{MathOverflow $T_l$}
    \end{subfigure}
    %\hfill
    \begin{subfigure}[b]{0.45\textwidth}
      \centering
    \includegraphics[width=\textwidth]{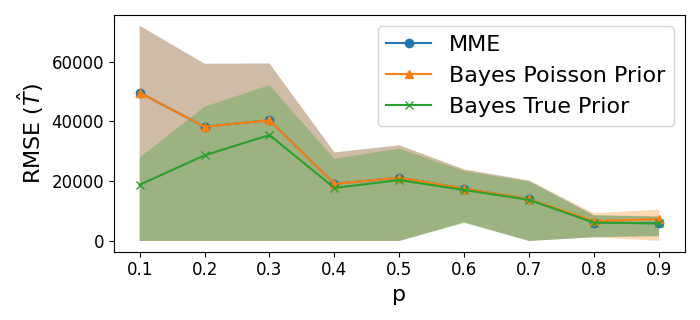}
    \caption{MathOverflow $T$}
    \end{subfigure}  
    \caption{Error in estimation of the triangles per link sequence (left hand column) and total triangles (right hand column) using our different approaches. Each value is averaged over 10 experiments with the shaded error bars representing standard deviation. The MME overlays the Bayes estimator with Poisson prior in the right hand column.}
    \label{fig:triangle_reconstruction}
\end{figure}

In the triangle count experiment, we estimate the triangle per edge count $\hat{T}_l$ for edges $e_l \in E'$ and compute the mean squared error as $\text{RMSE}(\hat{\mathbf{T}}) = \left[\frac{1}{M'}\sum_{e_l \in E'} (T_l - \hat{T}_l)^2\right]^{\frac{1}{2}}$. In addition, we estimate the total number of triangles as described in~\cref{sec:estimators} and calculate the mean squared error over the 10 experiments performed. These are shown in figure~\ref{fig:triangle_reconstruction}.
In all experiments, the Bayes estimator with Poisson prior overlays the MME for the total number of triangles because the $\lambda$ used in the Poisson distribution is the MME estimate of the average number of triangles per link.. However, in all but the AS topology, the Poisson prior improves the estimate of triangles per link, especially in the small $p$ scenario. In the AS topology dataset, the Poisson is an inappropriate prior, performing poorly even with large sample sizes.

\begin{figure}[htbp]
    \centering
    \begin{subfigure}[b]{0.47\textwidth}
    \centering
    \includegraphics[width=\textwidth]{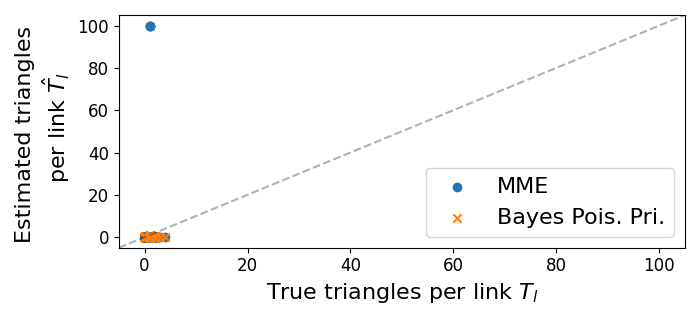}
    \caption{Erd\H{o}s-R\'enyi $p=0.1$}
    \end{subfigure}
    \begin{subfigure}[b]{0.47\textwidth}
    \centering
    \includegraphics[width=\textwidth]{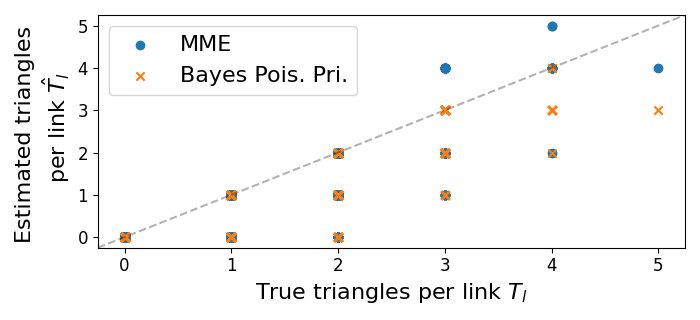}
    \caption{Erd\H{o}s-R\'enyi $p=0.9$}
    \end{subfigure}
    % \vskip\baselineskip
    \begin{subfigure}[b]{0.47\textwidth}
    \centering
    \includegraphics[width=\textwidth]{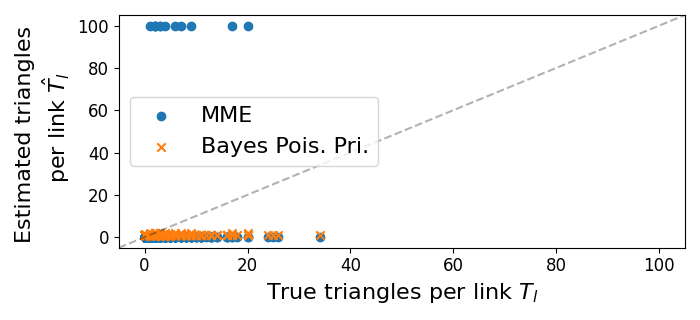}
    \caption{Barab\'asi-Albert $p=0.1$}
    \end{subfigure}
    \begin{subfigure}[b]{0.47\textwidth}
    \centering
    \includegraphics[width=\textwidth]{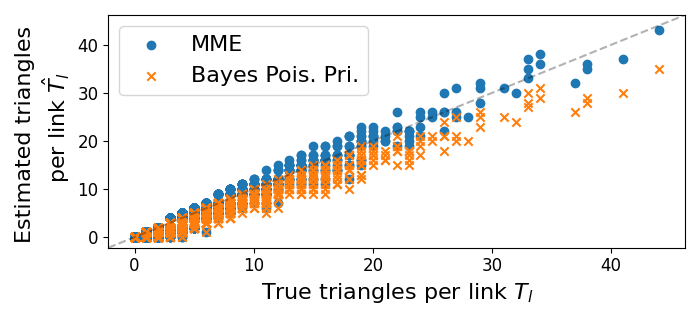}
    \caption{Barab\'asi-Albert $p=0.9$}
    \end{subfigure} 
    % \vskip\baselineskip
    \begin{subfigure}[b]{0.47\textwidth}
    \centering
    \includegraphics[width=\textwidth]{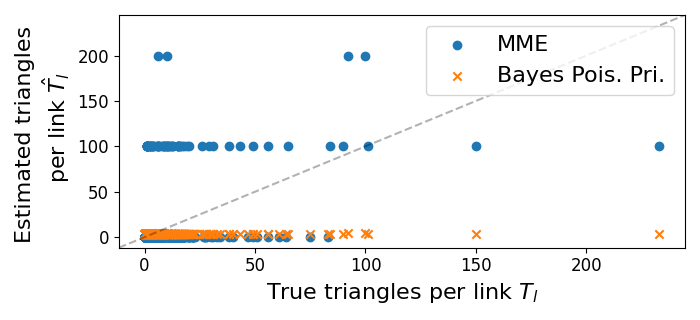}
    \caption{AS topology $p=0.1$}
    \end{subfigure}
    \begin{subfigure}[b]{0.47\textwidth}
    \centering
    \includegraphics[width=\textwidth]{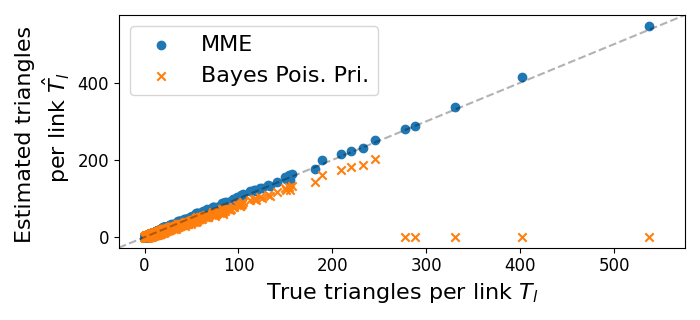}
    \caption{AS topology $p=0.9$}
    \end{subfigure} 
    % \vskip\baselineskip
    \begin{subfigure}[b]{0.47\textwidth}
    \centering
    \includegraphics[width=\textwidth]{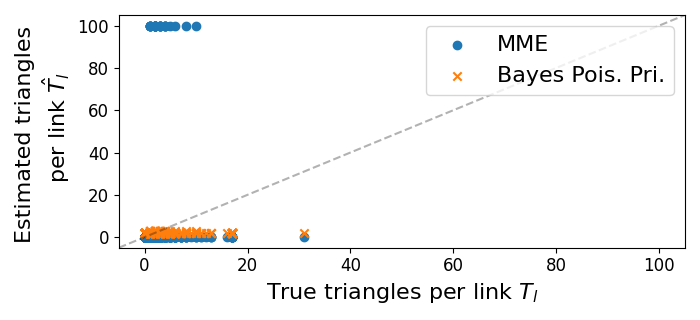}
    \caption{ArXiv Hep-Th $p=0.1$}
    \end{subfigure}
    \begin{subfigure}[b]{0.47\textwidth}
    \centering
    \includegraphics[width=\textwidth]{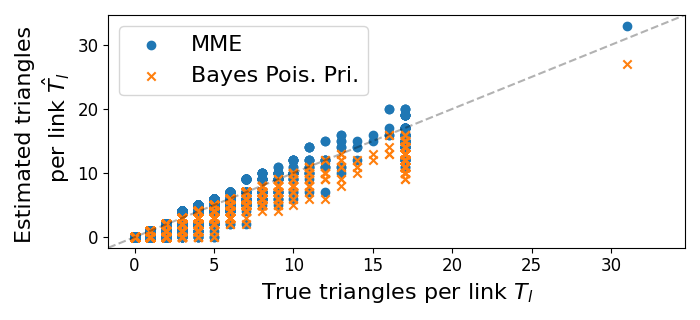}
    \caption{ArXiv Hep-Th $p=0.9$}
    \end{subfigure}
    % \vskip\baselineskip
    \begin{subfigure}[b]{0.47\textwidth}
    \centering
    \includegraphics[width=\textwidth]{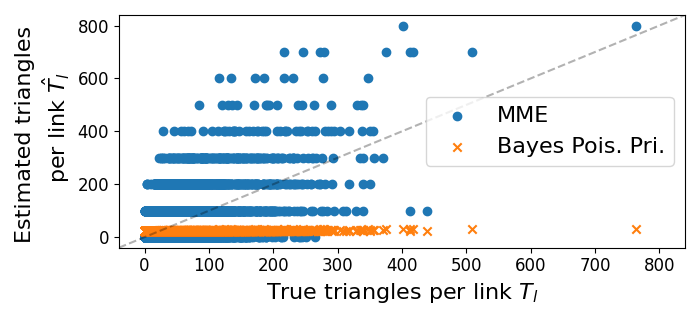}
    \caption{MathOverflow $p=0.1$}
    \end{subfigure}
    \begin{subfigure}[b]{0.47\textwidth}
    \centering
    \includegraphics[width=\textwidth]{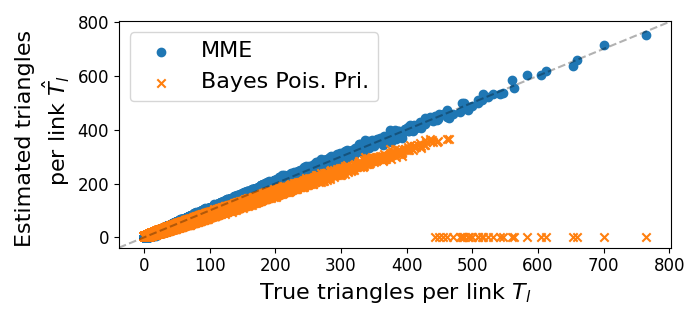}
    \caption{MathOverflow $p=0.9$}
    \end{subfigure}

    \caption{Estimated edge triangle count $\hat{T_l}$ against the true edge triangle count $T_l$ for five different datasets, at a $p=0.1$ sample and $p=0.9$ sample. We compare the method of moments estimator with the Bayes posterior estimate with Poisson prior.}
    \label{fig:tri_vs_estimated}
\end{figure}

As with the degree sequence, we compare the number of triangles per link and estimated triangles per link in small and large $p$ scenarios in our datasets, using the MME and Bayes posterior estimate with Poisson prior in figure~\ref{fig:tri_vs_estimated}. Considering first the $p=0.1$ scenario, we see that the Bayes posterior estimate helps to ease the problem of extreme estimates (the MME $\hat{T}_l = T'/p^2$ takes smallest values 0 then $1/p^2$) since the contribution of high $t$ terms in~\cref{eqn:posteriortriangles} is suppressed by the Poisson prior. However, for large $p$, we see that the Poisson prior begins to dominate inappropriately, incorrectly assuming that the distribution of triangles across the links is homogeneous. This manifests in the BA and Hep-Ph examples as a slight overestimate at the lower triangle count end and a large underestimate at the upper end. We see this even clearer in the AS topology example, where a triangle count of over 250 is effectively forbidden by the Poisson prior, while the MME remains unbiased in all datasets.
The poor performance of both estimators at $p=0.1$ indicates how difficult a problem estimating the edge triangle count is after so much information has been lost; in all datasets apart from the AS topology, edges in the sampled graph had only 0 or 1 triangle remaining. A better estimate for these may involve more assumptions about the network structure. For example, the results in figure ~\ref{fig:tri_vs_estimated} show that the distribution of triangles in the real network and, to an extent, the BA network is heterogeneous, so perhaps a heavier tailed distribution than a Poisson should be considered.
\section{Conclusion}\label{sec:conclusion}This paper provided methods for recovering the degree sequence, number of triangles and triangle per link sequence from networks sampled 
%in a canonical way -- I don't know why this phrase is here
via uniform edge sampling such as graphs limited to a sample by the Twitter API. %. This form of edge sampling is not only natural but a common restriction for datasets collected via APIs such as the Twitter API.
Our derivations of the expectation and variance of these quantities demonstrate the difficulty of the problem, finding their dependency on the degree sequence, degree-degree correlations and triangle distributions of the original networks.
Our results show that our Bayesian estimators perform much better than standard approaches on the degree sequence even when the priors were constructed without knowledge of distributions for the original network. For the triangle count per edge, we showed that while the Bayes estimates do not always improve upon the MME for total triangle counts, they provide a markedly better estimate of triangles per link in the small $p$ scenario where most information has been lost. However, an inappropriate choice of prior can lead to a bias even when the sample size is large, as we found when using a Poisson prior for a network with heterogeneous triangle distribution. 

Future work will investigate generalising methods we used for constructing a degree distribution prior for constructing a prior for triangle counts per link. However, it should be noted that, in particular, reconstructing triangles is extremely hard at low sampling rate and it is doubtful any technique would produce large improvements. The work could instead be extended by considering  sampling regimes and network properties for which a likelihood can be calculated.
\section*{Declarations}
\subsubsection*{Ethical approval}
No ethical approval is applicable for this work.
\subsubsection*{Competing interests}
The authors have no competing interests as defined by Springer, or other interests that might be perceived to influence the results and/or discussion reported in this paper.
\subsubsection*{Authors' contributions}
All authors designed the research and contributed equally to the manuscript.
\subsubsection*{Funding}
This work is part of the project ``Raphtory: a practical system for the analysis of dynamic graphs'' funded by the Alan Turing Institute.
\subsubsection*{Availability of data and materials} 
All real datasets are publicly available (referenced in the main text). Code for generating the synthetic networks and performing all experiments in the text is available at~\url{https://github.com/narnolddd/sample-nets}.
\bibliographystyle{abbrv}
\bibliography{biblio}
% \appendix
%  \section{Appendix}\label{sec:appendix}
%  \input{sections/appendix}
\end{document}